\documentclass[twocolumn]{aastex701}
\usepackage{amsmath}
\hypersetup{colorlinks=true,linkcolor=xlinkcolor,citecolor=xlinkcolor,urlcolor=xlinkcolor}
\defcitealias{2016MNRAS.459.3738I}{IHO16}

\begin{document}
\title{A Similarity Theorem and Its Breakdown in Atomic Black Hole Accretion}
\shorttitle{Similarity and Its Breakdown}
\shortauthors{DuPont}

\author[orcid=0000-0003-3356-880X,gname=Marcus,sname=DuPont]{Marcus DuPont}
\affiliation{Department of Astrophysical Sciences, Princeton University,
Princeton, NJ 08540, USA}
\email[show]{marcus.dupont@princeton.edu}

\begin{abstract}
Inferring black-hole mass and growth from radiation processed by dense
atomic gas requires connecting the emitting atmosphere to its feeding flow.
We derive an exact similarity for point-mass accretion and use its limits
to identify spectral information and dynamical consistency tests.
At fixed temperature and composition, scaling mass, radius, and time
together while scaling density inversely preserves the dimensionless flow,
optical depths, and normalized spectra within a specified class of
two-body microphysics, radiation, and boundary laws. Luminosity scales
with mass, and the full atomic cooling curve is retained. Within this
class the fixed-velocity power-law rescaling is unique.
Density-sensitive populations and thermal radiation break the symmetry;
public models of Little Red Dots illustrate the resulting density information.
Even with luminosity and density normalization known, a prescribed
spherical inflow retains an exact degeneracy between central mass and
inward drift. A proposed feeding model nevertheless supplies growth and
inventory tests. For point-mass Bondi feeding at fixed ambient sound speed,
luminosity and its conversion to retained growth determine the ratio of
adjustment to growth time independently of black-hole mass. For the
critical isothermal Bondi profile with full retention, comparable times
require gas mass comparable to the hole's mass. More generally, steady
spherical continuity links luminosity and gas transit time to a required
reservoir mass and minimum radial column. Lower retention slows growth
while leaving this inventory unchanged at fixed supply and transit time.
\par
\end{abstract}


\section{Introduction}\label{sec:intro}

The Little Red Dots discovered by the James Webb Space Telescope
(JWST) have motivated models in which dense gas processes the radiation
from an accreting black hole
\citep{2024ApJ...963..129M,2024ApJ...964...39G,2026Natur.649..574R}.
Their spectra carry information about the emitting gas. Recovering the
central mass and its growth also requires a model of the feeding flow.
We develop an atomic accretion similarity and use its limits to connect
spectral information to the dynamics and gas supply of these systems.

The Bondi problem has a familiar economy. For a central point mass
$M$ in gas with ambient sound speed $c_\infty$, the natural length is
the Bondi radius $R_{\rm B}$. We write
$t_{\rm B}=R_{\rm B}/c_\infty$ for its sound-crossing time.
Once radius is measured in $R_{\rm B}$,
velocity in $c_\infty$, and density in its ambient value $\rho_\infty$,
accretors of different masses share the same dimensionless flow \citep{1952MNRAS.112..195B}.

Atomic cooling and radiative transfer introduce clocks and columns that
could break this similarity. At fixed temperature, however, the two-body
cooling time is inversely proportional to density. Increasing the mass,
radii, and flow times while reducing density by the same factor preserves
the number of cooling times per crossing. It also preserves columns:
the lower density compensates for the longer path. With compatible
radiation and boundary laws, this rescaling retains the full atomic
cooling curve and gives complete similarity in the sense of
\citet{1996sssi.book.....B}.

Earlier radiative-accretion calculations found radii and cycle periods
proportional to $M$ and invariant Eddington-scaled rates along
fixed $n_\infty M$, where $n_\infty$ is the ambient number density
\citep{2011ApJ...739....2P,2012ApJ...747....9P,2012ApJ...754..154M}.
\citet[hereafter IHO16]{2016MNRAS.459.3738I} identified the same product as the
hyper-Eddington control parameter, with a luminosity-dependent boundary
also explored by \citet{2016MNRAS.461.4496S}. We establish the corresponding
exact time-dependent transformation and its uniqueness within a specified
class of power-law rescalings, then examine how additional microphysics,
self-gravity, and retained growth change the dimensionless problem.

For Little Red Dots, thermal-atmosphere and reprocessing-cocoon
calculations provide concrete spectral models
\citep{Liu2025drrkn,Liu2026mbxlp,Sneppen2026dcskp}.
The atmosphere models illustrate how density-sensitive spectra can
distinguish emitting conditions. Connecting those conditions to a central
mass requires dynamics: even at fixed luminosity and density profile,
a prescribed spherical inflow retains a degeneracy between mass and
inward drift.

A proposed conversion from luminosity to supply nevertheless
allows two physical tests. Can the gas adjust while the hole grows, and
does the feeding region contain enough gas to sustain the supply?
For point-mass Bondi feeding at fixed ambient sound speed, the ratio of
adjustment to growth time is independent of the assigned black-hole mass
at fixed luminosity, radiative efficiency, and retained fraction.
Steady spherical continuity, without distributed mass sources or sinks,
also relates supply and transit time to the required gas inventory and
a minimum radial column. These tests connect
the spectral interpretation to a flow capable of supporting it.

Sections~\ref{sec:theorem} and \ref{sec:consequences} develop the similarity
and its consequences. Section~\ref{sec:lrdfeeding} examines spectral
information and mass inference; Sections~\ref{sec:breaking} and
\ref{sec:inventory} derive the growth and gas-inventory tests.
Section~\ref{sec:conclusions} draws the conclusions.
Appendices~\ref{sec:growingderivations} and \ref{sec:lrdmethods} give
additional growth derivations and the public-data methods.

\section{The similarity transformation}\label{sec:theorem}

We use geometrized units with $G=c=1$, restoring these constants only when
evaluating dimensional quantities. Entropy per particle is dimensionless, and
$R_{\rm B}=M/c_\infty^2$.

Three physical requirements determine the scaling. Keeping the
gravitational velocity scale fixed makes radius proportional to mass;
keeping velocity fixed makes time proportional to radius; and keeping
the cooling time proportional to the flow time makes density inversely
proportional to mass. The conservation equations below establish that
this rescaling carries the entire solution, including its chemistry and
radiation when their terms have the corresponding weights.

Let $x_i$ be the Cartesian components of the position vector $\boldsymbol x$,
$r=(x_ix_i)^{1/2}$ the spherical radius, and $\partial_i\equiv\partial/\partial
x_i$. Let $\rho$, $v_i$, and $P$ denote the
instantaneous density, velocity, and gas pressure, and let $\mathcal V_{ij}$
be the collisional viscous stress. Repeated spatial indices are summed, and
$\delta_{ij}$ is the Kronecker delta. The point-mass potential is
$\Phi=-M/r$. Let $\mathcal E_{\rm int}$ be the internal gas energy
per unit volume and $\mathcal E$ the sum of internal and kinetic gas energy
per unit volume,
$\mathcal F_{E,i}$ its flux, $f_{{\rm rad},i}$ the radiative force per unit
volume, $H_{\rm rad}$ the radiative heating rate per unit volume,
$n_b=\rho/m_p$ the baryon number density, $m_p$
the proton mass, $T$ the temperature, $\Lambda_b(T)$ the two-body atomic
cooling coefficient, and $S_{\rm fb}$ any additional non-radiative feedback
power deposited per unit volume. Representative conservation equations are
\begin{equation}
 \partial_t\rho+\partial_i(\rho v_i)=0,
 \label{eq:mass}
\end{equation}
\begin{equation}
 \partial_t(\rho v_i)+
 \partial_j\!\left(\rho v_iv_j+P\delta_{ij}-\mathcal V_{ij}\right)
 =-\rho\partial_i\Phi+f_{{\rm rad},i},
 \label{eq:momentum}
\end{equation}
\begin{equation}
 \partial_t\mathcal E+\partial_i\mathcal F_{E,i}
 =-\rho v_i\partial_i\Phi-n_b^2\Lambda_b(T)+S_{\rm fb}+H_{\rm rad}.
 \label{eq:energy}
\end{equation}

We say that a source, flux, or boundary law \emph{respects the similarity} if
it scales in the same way as the other terms in its conservation equation.
The similarity applies when the gravitational field
is dominated by a point mass; the ambient sound speed and composition are
fixed; material cooling and chemistry are two-body; dimensional boundaries
scale with $R_{\rm B}$; and all remaining closure, transfer, and feedback
laws respect the similarity. Each closure must give the transformed
source or flux the scaling required by its conservation equation.
The material equation of state must be homogeneous in density at fixed
temperature and composition: $P/\rho$ and
$\mathcal E_{\rm int}/\rho$ depend only on $T$ and the species abundances.
Dilute ideal-gas mixtures satisfy this condition, including mixtures whose
composition evolves. Spherical symmetry is unnecessary:
rotation, multidimensional structure, and nonspherical boundaries are carried
along when their dimensionless forms are fixed.

The scaling is easiest to find by beginning with gravity. Preserving
$M/r$ while changing the central mass by a positive factor
$\lambda$ requires $r\mapsto\lambda r$. Crossing times then require
$t\mapsto\lambda t$, while velocities and temperatures remain fixed. Suppose
$\rho\mapsto\lambda^{w_\rho}\rho$, where $w_\rho$ is initially unknown. Hydrodynamic
energy terms scale as $\lambda^{w_\rho-1}$, whereas two-body cooling scales as
$\lambda^{2w_\rho}$. Synchronizing the clocks gives $w_\rho-1=2w_\rho$, and hence $w_\rho=-1$.

Consider the transformation
\begin{equation}
 \begin{aligned}
 M'&=\lambda M,& x_i'&=\lambda x_i,& t'&=\lambda t,\\
 \rho'&=\lambda^{-1}\rho,& P'&=\lambda^{-1}P,&
 \mathcal V_{ij}'&=\lambda^{-1}\mathcal V_{ij},\\
 v_i'&=v_i,& T'&=T
 \end{aligned}
 \label{eq:dilation}
\end{equation}
Primed fields on the left are evaluated at $(x_i',t')$ and unprimed fields on
the right at the corresponding point
$(x_i'/\lambda,t'/\lambda)$.

\textit{Atomic point-mass similarity.---}
Suppose the initial and boundary data in
equations~(\ref{eq:mass})--(\ref{eq:energy}) respect the similarity. Suppose
also that every material energy or momentum source per unit volume gains a factor
$\lambda^{-2}$, while every stress and energy flux gains a factor
$\lambda^{-1}$. Then equation~(\ref{eq:dilation}) maps every smooth solution for mass
$M$ to a solution for mass $\lambda M$. The result also holds
for flows containing shocks: each shock radius and time is multiplied by
$\lambda$.

\textit{Why it works.---}
Evaluate the original solution at
$(x_i'/\lambda,t'/\lambda)$. The chain rule supplies one factor
$\lambda^{-1}$ for every space or time derivative.  Both terms in continuity
therefore gain a factor $\lambda^{-2}$. Every inertial, pressure, and stress
term in momentum gains the same factor. Since the ratio $M/r$ is
unchanged, the potential is unchanged, its gradient gains a factor
$\lambda^{-1}$, and the gravitational force density gains
$\lambda^{-2}$. Material energy density and energy flux gain
$\lambda^{-1}$; hence their derivatives and, by hypothesis, every source gain
$\lambda^{-2}$. The transformed equations are exactly the original equations
evaluated at the corresponding point, multiplied by $\lambda^{-2}$.

For a flow containing discontinuities, apply the same change of variables to
the conservation laws integrated over a control volume. The surface fluxes
scale with the volume terms, so the usual mass, momentum, and energy jump
conditions across a shock are preserved. For an ideal gas of fixed
composition, write the entropy per particle as
$s=\mu m_p s_{\rm phys}$, where $s_{\rm phys}$ is entropy per unit mass and
$\mu$ is the mean molecular weight in proton-mass units. It transforms as
$s'=s+\ln\lambda$ everywhere. Its jump and the sign of entropy production are
therefore unchanged. More generally, if a numerical or physical
rule is used to select the physical shocked solution, that rule must itself
respect the similarity.

\textit{Uniqueness.---}
Among power-law rescalings that change $M$, preserve a nonzero
two-body cooling term, and leave temperature, composition, and characteristic
velocity fixed, equation~(\ref{eq:dilation}) is the only possibility (apart
from renaming the scale factor).

To see why, take $M\mapsto\lambda M$. Keeping
$M/r$ unchanged
requires $r\mapsto\lambda r$, and keeping velocity unchanged then requires
$t\mapsto\lambda t$. Suppose $\rho\mapsto\lambda^{w_\rho}\rho$. A material-energy
derivative scales as $\lambda^{w_\rho-1}$, whereas a nonzero two-body loss scales
as $\lambda^{2w_\rho}$. Thus $w_\rho-1=2w_\rho$ and $w_\rho=-1$. The equation of state then gives
$P,\mathcal E\mapsto\lambda^{-1}(P,\mathcal E)$, leaving no free exponent.

Nonequilibrium two-body chemistry rides along with the flow. For every species
$\mathsf S$, let $n_{\mathsf S}$ be its number density. It scales as
$n_{\mathsf S}\mapsto\lambda^{-1}n_{\mathsf S}$. A collisional
source $k_{\mathsf S;RT}(T)n_{\mathsf R}n_{\mathsf T}$, where $k_{\mathsf S;RT}$ is a two-body reaction-rate
coefficient and $\mathsf R$ and $\mathsf T$ label the reactants, therefore scales as
$\lambda^{-2}$, matching both
terms in the species continuity equation. A photoionization rate per particle
scales with the photon flux as $\lambda^{-1}$, so its rate per unit volume also
scales as $\lambda^{-2}$. Ionization fractions and time-dependent ionization
fronts are consequently preserved in reduced variables.

A central feedback law may be written $L=\eta(\dot m)\dot M_{\rm flow}$, using the
usual accretion-efficiency convention \citep{Frank2002mrrlp}, where $L$
is the luminosity, $\dot M_{\rm flow}$ the inward mass supply, $\eta$ the feedback efficiency,
$\dot m=\dot M_{\rm flow}/\dot M_{\rm Edd}$ the Eddington-scaled supply, and
$\dot M_{\rm Edd}=L_{\rm Edd}$ the Eddington accretion rate, with
$L_{\rm Edd}$ the Eddington luminosity. Both $\dot M_{\rm flow}$ and
$L_{\rm Edd}$ scale as $\lambda$, so $\dot m$ and $\eta(\dot m)$ are
invariant. The luminosity scales as $\lambda$ and, when deposited over a
similar volume, $S_{\rm fb}$ scales as $\lambda^{-2}$.

Eddington feedback also respects the similarity. For fixed
electron-scattering opacity $\kappa_{\rm es}$, $L_{\rm Edd}\propto M$, and the radiation
force density $\rho\kappa_{\rm es}F$ scales as $\lambda^{-2}$, where $F$ is the
radiation flux. A response delay
proportional to $M$ or to a local flow time is also carried along. A
fixed delay or an independently prescribed source or photospheric radius
introduces an additional scale.

Radiative transfer respects the similarity only when the emission and
absorption terms have the required scalings. The transfer equation itself
gives this condition. For
specific intensity $I_\nu$ at frequency $\nu$, take
$I_\nu\mapsto\lambda^{-1}I_\nu$ and leave frequency unchanged. In
the time-dependent transfer equation \citep[e.g.,][]{1979rpa..book.....R},
\begin{equation}
 \partial_t I_\nu+\hat n_i\partial_i I_\nu
 =j_\nu-\alpha_\nu I_\nu,
 \label{eq:transfer}
\end{equation}
where $\hat n_i$ is the photon propagation direction, $j_\nu$ is the
emissivity, and $\alpha_\nu$ is the extinction per unit length. Every term
gains a factor $\lambda^{-2}$ provided $j_\nu$ gains
$\lambda^{-2}$ and extinction per unit length $\alpha_\nu$ gains
$\lambda^{-1}$.
Two-body atomic emissivity and
$\alpha_\nu=\rho\kappa_\nu(T,Y_A)$ satisfy these conditions, where
$\kappa_\nu$ is the opacity per unit mass and $Y_A=n_A/n_b$ is the abundance
of species $A$. A central luminosity proportional to
$\lambda$
supplies the corresponding boundary condition.  Under these hypotheses the
column $\rho r$ and optical depth
$\tau_\nu=\int\kappa_\nu(T,Y_A)\rho\,dr$ are invariant. The radiation flux
scales as $L/r^2\mapsto\lambda^{-1}L/r^2$,
so absorption per unit volume, proportional to $\kappa_\nu\rho L/r^2$, scales as
$\lambda^{-2}$. The ionization parameter $L/(n_br^2)$ is invariant as well.
For an opacity linear in density, a local-thermodynamic-equilibrium (LTE) source
$j_\nu=\alpha_\nu B_\nu(T)$ gains only $\lambda^{-1}$ at fixed $T$, where
$B_\nu$ is the Planck function, and is a symmetry-breaking term. The specified
transfer class therefore includes two-body diffuse recombination emission and
leaves LTE emission outside the family.

Coherent electron scattering and Compton exchange also preserve the result:
their transfer and energy terms scale as $n_eI_\nu$ and
$n_eU_{\rm rad}\propto\lambda^{-2}$, where $n_e$ is the electron number
density and $U_{\rm rad}$ is the radiation energy density. An imposed radiation bath instead
introduces an external energy density.

\textit{Other transport and additional scales.---}
Each departure from the hypotheses supplies a physical scale. Collisional
viscosity and conduction remain inside the similarity family when their microscopic
cross sections depend only on the fixed thermodynamic state: the dynamic
viscosity and thermal conductivity then remain fixed while their flux
divergences scale as $\lambda^{-2}$. A prescribed fixed kinematic viscosity or
thermal diffusivity introduces a scale outside the family. Ideal magnetohydrodynamics respects
the transformation for magnetic field
$B\mapsto\lambda^{-1/2}B$.

Higher-body and externally scaled processes lie outside the similarity. More generally, a local process involving $N$
interacting particles scales as
$\lambda^{-N}$, whereas the time-dependent terms scale as $\lambda^{-2}$.
Its relative importance therefore changes as $\lambda^{2-N}$; a three-body
term becomes stronger toward the denser, smaller-$M$ end. LTE radiation
energy proportional to $T^4$, fixed microphysical diffusivity, a fixed outer
radius, an external variability time,
cosmological expansion, or an untransformed relative velocity likewise
introduces another scale \citep[e.g.,][]{2007ApJ...662...53R,2017PhRvD..95d3534A}.
Nonideal equations of state with an intrinsic density scale likewise lie
outside the family.

\section{What one solution represents}\label{sec:consequences}

Let $\dot M_{\rm B}$ denote the Bondi supply. The rescaling
leaves
\begin{equation}
 \frac{r}{R_{\rm B}},\quad \frac{t}{t_{\rm B}},\quad
 \frac{\rho}{\rho_\infty},\quad \frac{\dot M_{\rm flow}}{\dot M_{\rm B}},\quad
 \frac{\dot M_{\rm flow}}{\dot M_{\rm Edd}},\quad \frac{L}{L_{\rm Edd}},\quad \tau_\nu
 \label{eq:invariants}
\end{equation}
unchanged. At fixed temperature and composition, each similarity family lies on
\begin{equation}
 n_\infty M={\rm constant}.
 \label{eq:family}
\end{equation}
The dimensionless Bondi supply gives this coordinate a direct physical
meaning:
\begin{equation}
 \dot m_{\rm B}\equiv\frac{\dot M_{\rm B}}{L_{\rm Edd}}
 \propto\frac{\rho_\infty M}{c_\infty^3}.
 \label{eq:bondisupplycoordinate}
\end{equation}
At fixed ambient state, $\dot m_{\rm B}$ labels the similarity family and
measures the available supply in Eddington units. The equivalent atomic
cooling coordinate $t_{\rm B}/t_{{\rm cool},\infty}$, where
$t_{{\rm cool},\infty}$ is the ambient cooling time, is proportional to the
same product.

The converse requires the remaining dimensionless data to agree: spectra,
boundary shapes, chemical abundances, and feedback functions must also match.

\noindent\textit{Collapse test.}---
For a field $U$ with characteristic scale $U_0$, define
\begin{equation}
 \widehat U(\boldsymbol\xi,\mathcal N)
 \equiv\frac{U(R_{\rm B}\boldsymbol\xi,t_{\rm B}\mathcal N)}{U_0},
 \quad \boldsymbol\xi=\frac{\boldsymbol x}{R_{\rm B}},\quad
 \mathcal N=\frac{t}{t_{\rm B}}.
 \label{eq:reducedfield}
\end{equation}
Here $U_0$ carries the same similarity weight as $U$; examples are
$\rho_\infty$, $c_\infty$, and $L_{\rm Edd}$. Two transformed simulations
must give the same $\widehat U$ at every shared $(\boldsymbol\xi,\mathcal N)$.
Any residual structure then locates physical or numerical scales that were
left untransformed. When several branches coexist, initial and boundary data
select the pair being compared.
This use of similarity as a numerical benchmark follows the successful SPH
test of \citet{2000MNRAS.314..759A}, who recovered analytic shocked-cooling
profiles at successive times for a specially chosen power-law cooling
function. The invariant temperature here allows the same test with the full
atomic cooling curve.

Equation~(\ref{eq:family}) explains why radiative-accretion boundaries so often
trace $n_\infty\propto M^{-1}$. Conditions based on the ratio of an
ionized or cooling radius to $R_{\rm B}$, on optical depth, or on an
Eddington-scaled supply are level sets of the same invariant. Their
normalizations still depend on the spectrum, chemical network, and feedback
prescription. Their common slope follows before those details are chosen.

There is also a dimensional consequence. Let $\dot M$ denote the rate
actually retained by the black hole. If the retained fraction is unchanged
along the family, then $\dot M$ scales as $\lambda$ and the
instantaneous physical growth time
$t_{\rm grow}=M/\dot M$ is shared by the family. Dynamical
times, meanwhile, scale as $M$.
Larger and more rarefied members evolve more slowly in their local flow time,
yet require the same number of years to grow by a fixed factor.

Each accretion state has its own similarity family. Equation~(\ref{eq:family})
says which problems are equivalent; stability and boundary conditions select
the solution. Near the hyper-Eddington boundary, for example, spherical
calculations find episodic and steady branches as the ionized region crosses
$R_{\rm B}$ \citepalias{2016MNRAS.459.3738I}.

\section{Spectral information and black-hole mass}\label{sec:lrdfeeding}

The theorem gives a reference point for interpreting a spectrum: within
its transfer class, an entire family has the same normalized shape.
Density-sensitive departures can distinguish emitting atmospheres. We
first examine that information in public spectral models, then ask what
dynamics is needed to connect it to black-hole mass and retained growth.

\subsection{Spectral consequences of atomic similarity}
The atomic transformation sends $M,r,t$ to
$\lambda M,\lambda r,\lambda t$, while $\rho,P,I_\nu$ scale as $\lambda^{-1}$ and
temperature and velocity remain fixed. With corresponding boundaries and
the same source spectrum, integration over emitting area gives
\begin{equation}
 L_\nu' =\lambda L_\nu,\qquad
 \frac{L_\nu'}{L'}=\frac{L_\nu}{L}.
\end{equation}
Members of the family share their colors, equivalent widths, and the
shapes of corresponding velocity-broadened profiles. Their physical clocks
stretch by $\lambda$. This correspondence follows the same emission
components and geometry through the transformation. An observed variability
time becomes useful once its physical origin is identified. Absolute
luminosity provides another route: calibrating $L/M$ within a
family supplies its mass scale.

Thermal radiation offers an instructive comparison. For
density-independent mass opacity, fixed optical depth and temperature can
permit $r\rightarrow \lambda r$, $\rho\rightarrow\rho/\lambda$ with fixed intensity and
$L\rightarrow \lambda^2L$. Related radiative-transfer scalings are familiar from
dust emission \citep{Ivezic1996tssyj}. Holding the intensity fixed changes
how the radiation force scales, leading to a different dynamical balance.

The transfer equation shows where the two scalings separate. For local
$\alpha_\nu\propto\rho^{b_\alpha}$ and $j_\nu\propto\rho^{b_j}$ at fixed
thermodynamic state, let $b_\alpha$ and $b_j$ denote the density
exponents of absorption and emission. The atomic intensity transformation requires
\begin{equation}
 b_\alpha=1,\qquad b_j=2.
\end{equation}
In local thermodynamic equilibrium, Kirchhoff's law
$j_\nu=\alpha_\nu B_\nu(T)$ \citep{1979rpa..book.....R} requires the same density dependence in
emission and absorption. LTE therefore breaks the atomic transformation.
For example, free--free emission has the required quadratic emissivity,
while its quadratic true absorption changes the optical depth along the
family. A thermal photosphere
also obeys $L=4\pi R_{\rm ph}^2\sigma_{\rm SB} T_{\rm eff}^4$, where
$R_{\rm ph}$ is the emitting radius, $T_{\rm eff}$ its effective temperature,
and $\sigma_{\rm SB}$ the Stefan--Boltzmann constant. Imposing the atomic
$R_{\rm ph}\propto M$ and $L\propto M$ would require
$T_{\rm eff}\propto M^{-1/4}$. This temperature change moves the
photosphere outside the fixed-temperature atomic family. Here
$T_{\rm eff}$ is defined by the bolometric luminosity and emitting area;
inferring it from a continuum color requires a spectral model.

Departure from the symmetry can begin before LTE. The two-level atom
provides a simple example \citep{Osterbrock2006rderb}. Electrons excite ions
from the lower level $l$ to the upper level $u$; spontaneous radiation and
electron collisions return them to $l$. In statistical equilibrium these
rates balance:
\begin{equation}
 n_l n_e q_{lu}=n_u(A_{ul}+n_e q_{ul}).
 \label{eq:twolevelbalance}
\end{equation}
Here $n_l$ and $n_u$ are the level populations, $A_{ul}$ is the
spontaneous radiative rate, and $q_{lu}$ and
$q_{ul}$ are electron-impact excitation and de-excitation coefficients.
Taking the line to be optically thin and neglecting external radiative
pumping, we substitute $n_l=n_X-n_u$, where $n_X$ is the total ion density,
and solve for the upper population:
\begin{equation}
 n_u=\frac{n_Xn_eq_{lu}}{A_{ul}+n_e(q_{lu}+q_{ul})}.
\end{equation}
The denominator introduces the population turnover density
$n_*=A_{ul}/(q_{lu}+q_{ul})$. Keep the temperature, abundance fraction,
and electron fraction fixed along the atomic family, and write
$\zeta=n_{e,0}/n_*$. Both $n_X$ and $n_e$ then scale as $\lambda^{-1}$.
The emitted line power is $h\nu_{ul}A_{ul}n_u$ per unit volume, while a
corresponding volume scales as $\lambda^3$. Combining these factors gives
\begin{equation}
 \frac{L_{ul}(\lambda)}{\lambda L_{ul}(1)}=\frac{1+\zeta}{1+\zeta/\lambda}.
\end{equation}
The quadratic low-density limit recovers atomic similarity. Finite
collisional de-excitation introduces a density scale. Real line diagnostics
require the appropriate multilevel populations, ionization balance, and
escape calculation; the two-level example establishes the mechanism.

\subsection{Density information in public atmosphere models}
The public spherical models of \citet{Liu2025drrkn} prescribe
$\rho=Ar^{-3/2}$, with density normalization $A$, and independently
specify luminosity. We compare all
15 released spectra along five tracks with $A\propto L^{1/2}$ spanning
two decades in luminosity. These are the density--luminosity weights of
the atomic transformation for this radial power law. Their thermal
structures and boundaries also enter the comparison. The normalized
spectra change along these tracks, with much of the change absorbed by
a fitted amplitude and smooth continuum tilt: optical residuals on four
tracks are $0.010$--$0.019$ mag, while the densest track retains
larger departures. These comparisons illustrate how a spectral change
can trade against a smooth contaminating component. The residuals combine
opacity, temperature, and boundary effects; separating those contributions
requires the per-run radial states and boundary radii.

The atmosphere library of \citet{Liu2026mbxlp} supplies an independent
test of density discrimination. Its authors already identify continuum
curvature, the H$^-$ feature near $1.6\,\mu$m, and Ca-triplet absorption
as density diagnostics. The library's effective gravity $g$ describes the net
atmospheric support and serves primarily as a proxy for photospheric density.
We use its convention $\log g=\log_{10}[g/({\rm cm\,s^{-2}})]$.
We compare noiseless targets with all 242 radiation-force-on templates
using optical coverage and successive additions of rest-frame J, H, and K
bands. We repeat the comparison with host light, dust emission, and
extinction, assigning each band a 0.05-mag error scale. This controlled
experiment isolates how wavelength coverage and the adopted contaminating
components affect the range of compatible photospheric densities
$\rho_{\rm ph}$. Appendix~\ref{sec:lrdmethods} specifies the targets,
bands, component choices, fitting procedure, and numerical controls.

\begin{figure*}[t]
\centering
\includegraphics[width=\textwidth]{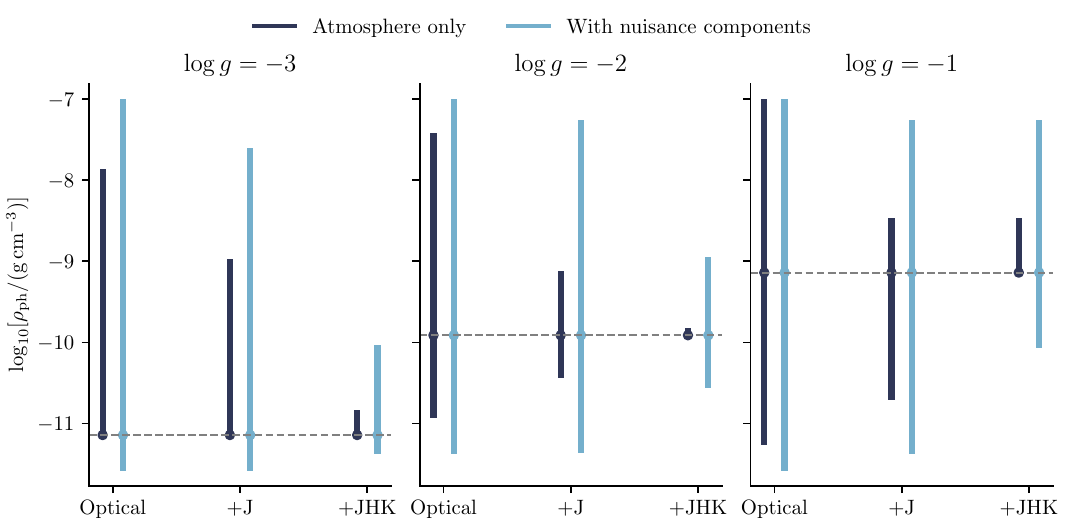}
\caption{Photospheric density information gained by extending wavelength
coverage in public atmosphere models. The panels show three noiseless
4500 K targets labeled by effective gravity,
$\log g=\log_{10}[g/({\rm cm\,s^{-2}})]$. The horizontal dashed lines
and points mark their photospheric densities $\rho_{\rm ph}$. Bars span
the densities of compatible templates, first using optical bands and then
adding rest-frame J and JHK. Compatibility means $\Delta\mathcal Q\leq4$,
where $\mathcal Q$ is the sum of squared flux residuals weighted by
assigned 0.05-mag errors, minimized over fitted component parameters.
Its increase above the best-fitting template is $\Delta\mathcal Q$.
The two colors compare an atmosphere alone with one including host
light, dust emission, and extinction. The widths measure discrimination
within the discrete library at this tolerance.}
\label{fig:lrddensity}
\end{figure*}

Let $\mathcal Q$ denote the weighted sum of squared flux residuals and
$\Delta\mathcal Q$ its increase above the best-fitting template after
minimizing over nuisance parameters (Appendix~\ref{sec:lrdmethods}).
Figure~\ref{fig:lrddensity} shows three 4500 K targets. For the
lowest-density target with contaminating emission and extinction,
the density range of templates satisfying $\Delta\mathcal Q\leq4$
narrows from 4.51 dex with optical coverage to 1.27 dex with JHK added.
The higher-density targets retain wider degeneracies. These widths
measure discrimination among the supplied atmospheres at the adopted
tolerance; the grid spacing and boundaries also affect them.

The experiments show how departures from similarity become useful
observables once a spectral model is specified. The resulting density
inference depends on that model and on the fitted contaminating
components. The partially reprocessing cocoons of \citet{Sneppen2026dcskp}
have an approximate spectral scaling of their own; including H$^-$ would
enable a closer comparison around the near-infrared feature. Connecting
any of these emitting atmospheres to a central mass then requires its
geometry and dynamical support.

\subsection{From density information to central mass}
Within an atomic similarity family, a calibrated $L/M$ lets
absolute luminosity set the mass scale. A prescribed atmosphere leaves
another freedom. Suppose radiation modeling determines both luminosity
and the full density profile. Holding these inputs fixed, we ask which
masses and flows can produce them. This differs from the atomic similarity,
along which luminosity scales with mass.

For the spherical profile $\rho=Ar^{-3/2}$ used by \citet{Liu2025drrkn}, prescribe
$|v_r|=\delta\sqrt{M/r}$. The dimensionless parameter $\delta$
sets the inward drift relative to the gravitational velocity scale.
Writing $L=\eta_{\rm rad}\dot M_{\rm flow}$ for the radiative efficiency
\citep{Frank2002mrrlp}, steady continuity
$\dot M_{\rm flow}=4\pi r^2\rho|v_r|$ gives
\begin{equation}
 A=\frac{\dot M_{\rm flow}}{4\pi\delta\sqrt{M}}.
\end{equation}
Even if radiation modeling determines both $L$ and $A$ exactly, four
parameters remain: $M,\dot M_{\rm flow},\delta,\eta_{\rm rad}$.
We can see how they trade against one another by taking logarithmic
differentials of the two relations:
\begin{equation}
 \begin{aligned}
 d\ln L&=d\ln\dot M_{\rm flow}+d\ln\eta_{\rm rad},\\
 d\ln A&=-\tfrac12d\ln M+d\ln\dot M_{\rm flow}
            -d\ln\delta.
 \end{aligned}
 \label{eq:lrdlogresponses}
\end{equation}
Each differential is a fractional change, $d\ln X=dX/X$.
The coefficients form a Jacobian: a matrix recording how the two
outputs respond to changes in the four inputs. Define
$\boldsymbol\theta=(\ln M,\ln\dot M_{\rm flow},\ln\delta,
\ln\eta_{\rm rad})$ and $\boldsymbol o=(\ln L,\ln A)$, with logarithms
taken relative to fixed reference units. Then
$d\boldsymbol o=J\,d\boldsymbol\theta$, where
\begin{equation}
 J_{ij}\equiv\frac{\partial o_i}{\partial\theta_j},\qquad
 J=\begin{pmatrix}0&1&0&1\\-1/2&1&-1&0\end{pmatrix}.
 \label{eq:lrdjacobian}
\end{equation}
A parameter displacement $\boldsymbol w$ satisfying
$J\boldsymbol w=0$ leaves both outputs unchanged; such displacements form
the matrix's null space \citep{Golub1996iesky}. The two independent rows
give rank two, leaving two independent directions in the four-dimensional
parameter space:
\begin{equation}
 \boldsymbol w_1=(1,0,-1/2,0),\qquad
 \boldsymbol w_2=(0,1,1,-1).
\end{equation}
Along the first direction, increasing $M$ by a factor of four and
halving $\delta$ preserves $L$, $A$, and the prescribed radial velocity at
each radius. Along the second, multiplying $\dot M_{\rm flow}$ and
$\delta$ by the same factor while dividing $\eta_{\rm rad}$ by that
factor preserves $L$ and $A$. These finite rescalings are exact because
the model relations are power laws. They apply over the parameter range
allowed by the adopted physical assumptions.
The prescribed spectral model therefore admits a continuous family of
mass and flow interpretations. The momentum and energy equations supply
the next constraints: solving them would determine which members can
sustain the required flow.

If both drift and radiative efficiency are supplied independently, then
\begin{equation}
 M=
 \left(\frac{L}{4\pi\eta_{\rm rad}\delta A}\right)^2.
\end{equation}
The mass depends quadratically on the adopted efficiency and drift.
A density measurement becomes a mass estimate through these dynamical
ingredients. Resolved gravitational kinematics, a calibrated dynamical
clock, or a model of atmospheric support can supply the additional information.

Even while this mass freedom remains, a proposed luminosity-to-supply
conversion specifies the gas passing through the system. Its retained
fraction sets the growth rate, and its transit time sets the reservoir
mass. The following sections use these quantities to test whether a
proposed interpretation can sustain its feeding flow.

\section{Growth and the response of the feeding gas}\label{sec:breaking}

A spectral interpretation specifies a radiating state. To use its
inferred supply for black-hole growth, the feeding gas must also respond
to the changing central potential. We first derive the growth criterion
for the point-mass family, then express it through luminosity and the
response time of a specified feeding model.

\subsection{When black-hole growth becomes dynamical}

\begin{figure}
 \centering
 \includegraphics[width=\columnwidth]{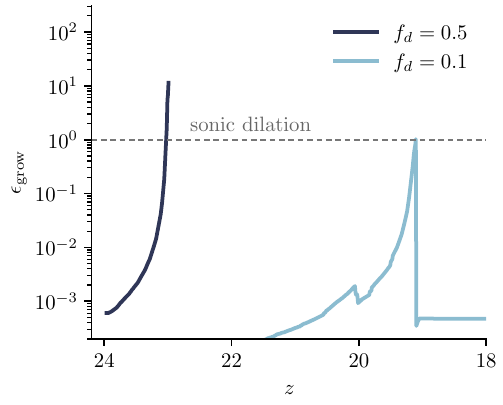}
 \caption{Growth loading along the published quasar histories of
 \citet{2005ApJ...633..624V}. The horizontal axis is redshift $z$, decreasing
 toward later cosmic time; the vertical axis is the retained mass gained per
 Bondi time, $\epsilon_{\rm grow}$. We digitized their public vector Figure~1 and
 combined its black-hole mass and Eddington-scaled accretion-rate tracks at
 the model temperature $T=8000\,{\rm K}$. The disk gas fraction $f_d=0.5$
 crosses the sonic-dilation boundary; $f_d=0.1$ reaches order
 unity as its supercritical episode ends. The abrupt endings belong to the
 original episodic histories.}
 \label{fig:vrgrowth}
\end{figure}

The symmetry maps flows around black holes of fixed mass. This is also how
the published hyper-Eddington solutions are constructed: $M$ sets a
prescribed potential and a fixed domain in units of its Bondi radius, while
the resulting accretion rate is later used to infer growth
\citepalias{2016MNRAS.459.3738I}. The coupled growth
equation $dM/dt=\dot M$ has a different scaling. Under
$M'(t')=\lambda M(t'/\lambda)$, the left-hand side is
unchanged, whereas the retained rate in the transformed flow is larger by
$\lambda$. Black-hole growth therefore does not share the symmetry.

This mismatch has a direct physical measure. Define
\begin{equation}
 \epsilon_{\rm grow}\equiv\frac{t_{\rm B}}{t_{\rm grow}}
 =\frac{\dot M t_{\rm B}}{M},
 \label{eq:growthloading}
\end{equation}
the instantaneous fractional growth rate measured per Bondi time. Although the
instantaneous growth time $t_{\rm grow}$ is the same for all
fixed-mass members, $t_{\rm B}\propto M$, and hence
$\epsilon_{\rm grow}'=\lambda\epsilon_{\rm grow}$. Writing the retained Eddington-scaled rate as
$\dot m_{\rm ret}\equiv\dot M/L_{\rm Edd}$ gives
\begin{equation}
 \epsilon_{\rm grow}=
 \frac{4\pi M\dot m_{\rm ret}}
 {\kappa_{\rm es}c_\infty^3}.
 \label{eq:growthloadingphysical}
\end{equation}
Restoring constants and normalizing the ambient temperature, mean molecular
weight in proton-mass units, and opacity to $T_\infty=10^4\,{\rm K}$, $\mu=1.22$, and
$\kappa_{\rm es}=0.4\,{\rm cm^2\,g^{-1}}$ gives the expression below.
The temperature and molecular weight describe a neutral atomic feeding
medium; the opacity fixes the Eddington-rate normalization. We choose
$10^4M_\odot$ and $\dot m_{\rm ret}=5000$ as reference values to connect
with the hyper-Eddington calculations of \citetalias{2016MNRAS.459.3738I}.
\begin{equation}
 \begin{aligned}
 \epsilon_{\rm grow}\simeq{}&0.834
 \left(\frac{M}{10^4M_\odot}\right)
 \left(\frac{\dot m_{\rm ret}}{5000}\right)
 \left(\frac{T_\infty}{10^4\,{\rm K}}\right)^{-3/2}\\
 &\times\left(\frac{\mu}{1.22}\right)^{3/2}
 \left(\frac{\kappa_{\rm es}}{0.4\,{\rm cm^2\,g^{-1}}}\right)^{-1}.
 \end{aligned}
 \label{eq:growthnumber}
\end{equation}
The sonic-growth boundary, $\epsilon_{\rm grow}=1$, is therefore
\begin{equation}
 \begin{aligned}
 M\dot m_{\rm ret}\simeq{}&6.00\times10^7M_\odot
 \left(\frac{T_\infty}{10^4\,{\rm K}}\right)^{3/2}
 \left(\frac{\mu}{1.22}\right)^{-3/2}\\
 &\times\left(\frac{\kappa_{\rm es}}
 {0.4\,{\rm cm^2\,g^{-1}}}\right).
 \end{aligned}
 \label{eq:sonicgrowthboundary}
\end{equation}
This boundary reaches beyond the hyper-Eddington regime. At the fiducial gas
temperature, a retained $\dot m_{\rm ret}=100$ track crosses it near
$M=6.00\times10^5M_\odot$, while a retained
$\dot m_{\rm ret}=1000$ track crosses near $6.00\times10^4M_\odot$.
Steady hyper-Eddington inflow requires $\dot m\gtrsim5000$, and the published
$10^4M_\odot$, $10^5\,{\rm cm^{-3}}$ solution reaches
$\dot m\simeq8000$ \citepalias{2016MNRAS.459.3738I}. If this inflow is
retained, equation~(\ref{eq:growthnumber}) gives
$t_{\rm B}\simeq7.56\times10^4\,{\rm yr}$,
$t_{\rm grow}\simeq5.66\times10^4\,{\rm yr}$, and
$\epsilon_{\rm grow}\simeq1.33$: at this retained rate, the hole gains
its original mass in less than one Bondi crossing. Both timescales are familiar; their ratio
also measures the speed of a boundary that was fixed in constructing the
solution. Because $R_{\rm B}=M/c_\infty^2$,
\begin{equation}
 \frac{\dot R_{\rm B}}{c_\infty}
 =\frac{\dot M t_{\rm B}}{M}
 =\epsilon_{\rm grow}.
 \label{eq:bondiexpansionspeed}
\end{equation}
Thus $\epsilon_{\rm grow}$ is both a growth-to-flow ratio and the dimensionless
dilation rate of the capture scale. At unity, the potential changes by order
unity in one global flow time and the fixed-mass continuation loses its
adiabatic small parameter. The Bondi radius measures the scale at which
the hole's gravity competes with ambient thermal motion; its expansion
tracks the changing reach of that gravity.

Here the retained rate matters. A super-Eddington inflow at the Bondi radius
may lose most of its mass to winds before reaching the hole. Equations
(\ref{eq:growthloading})--(\ref{eq:sonicgrowthboundary}) use the growth rate
that actually changes $M$. The retained fraction therefore
controls how quickly the supplied gas expands the hole's capture scale.

The same distinction bears on population arguments. Integrating the quasar
luminosity function constrains the secular mass growth density
\citep{1982MNRAS.200..115S}, while individual black holes grow during active
episodes. If $f_{\rm duty}$ is the duty cycle and $\dot M_{\rm on}$ is the
mean retained rate conditional on activity, the active and secular loadings obey
\begin{equation}
 \langle\dot M\rangle
 =f_{\rm duty}\dot M_{\rm on},\qquad
 \epsilon_{\rm on}
 =\frac{\dot M_{\rm on}}{c_\infty^3}
 =\frac{\epsilon_{\rm sec}}{f_{\rm duty}},
 \label{eq:dutyloading}
\end{equation}
where $\epsilon_{\rm sec}\equiv
\langle\dot M\rangle/c_\infty^3$. Thus a population can assemble
gently in the mean while its active members cross the sonic-dilation boundary.
Published accretion histories can therefore be translated into
$\epsilon_{\rm on}(z)$ once the ambient sound speed and retained active rate
are specified. Equation~(\ref{eq:dutyloading}) also makes clear that a
Soltan-style mean requires an episode duration and duty cycle to assess
how much time an individual flow has to relax between episodes.

Figure~\ref{fig:vrgrowth} makes this translation for the early supercritical
quasar tracks of \citet{2005ApJ...633..624V}. Their atomic disk has
$T=8000\,{\rm K}$. Restoring $c$, their Eddington accretion-rate normalization
is $L_{\rm Edd}/c^2$. Under their assumption that the
black hole accepts most of the inflow, the fuel-rich $f_d=0.5$ track reaches
$\epsilon_{\rm grow}\simeq11$ near $z=23$; the $f_d=0.1$ track grazes unity
near $z=19$. Here $f_d$ is the fraction of halo gas that settles into their
central isothermal disk. Expressing the published mass and accretion-rate
tracks in the growth loading reveals when the Bondi scale expands at the
ambient sound speed.

The growth loading also limits the flow time available during a
sustained episode. For a retained law $\dot M\propto M^p$
with $p>0$, let $\epsilon_0$ be the loading at the initial mass $M_0$.
A mass increase by a factor $f_M=M/M_0$ accumulates
$(1-f_M^{-p})/(p\epsilon_0)$ instantaneous Bondi times. At
the illustrative $\epsilon_0=1.3$, a mass doubling therefore allows only
$0.385$ Bondi times for $p=1$ and $0.288$ for $p=2$. An indefinitely sustained law has
the finite limit $1/(p\epsilon_0)$. These are conditional clock budgets;
the reservoir and feeding conditions determine how long the law applies.

Allowing the capture radius to grow also changes the reduced flow
equations. With a fixed retained fraction and nonzero accretion flux, a
stationary profile in these coordinates cannot preserve the original
atomic density scaling. The flow must acquire new time dependence or
depart from that scaling. Appendix~\ref{sec:growingderivations} derives
both this obstruction and the accumulated-clock result.

A fixed-mass calculation still gives an instantaneous member of the
similarity family. Its quasi-steady evolutionary interpretation requires
$\epsilon_{\rm grow}\ll1$. The analysis therefore identifies a concrete
regime in which the central mass and the expanding capture region must be
evolved together with the flow.

\subsection{Testing a luminosity-based growth interpretation}\label{sec:lumgrowth}
Even when the spectrum leaves the central mass uncertain, an accretion
interpretation assigns a mass supply, radiative efficiency, and retained
fraction to the measured luminosity. These choices imply a growth rate
that can be compared with the response time of the feeding gas.
Write $L=\eta_{\rm rad}\dot M_{\rm flow}$ and
$\dot M=f_{\rm ret}\dot M_{\rm flow}$, so the radiative efficiency
$\eta_{\rm rad}$ and retained fraction $f_{\rm ret}$ refer to the same
inward mass supply $\dot M_{\rm flow}$.

The feeding model must specify which gas responds to the growing hole
and what sets its adjustment time. A Bondi sound-crossing time applies
when the hole dominates gravity throughout the Bondi-scale feeding
region and the adopted ambient sound speed describes that gas. A compact
emitting photosphere alone does not establish these conditions. When
the host contributes appreciably to gravity, the gas model must supply
its own response time.

\begin{figure}[!tbp]
\centering
\includegraphics[width=\columnwidth]{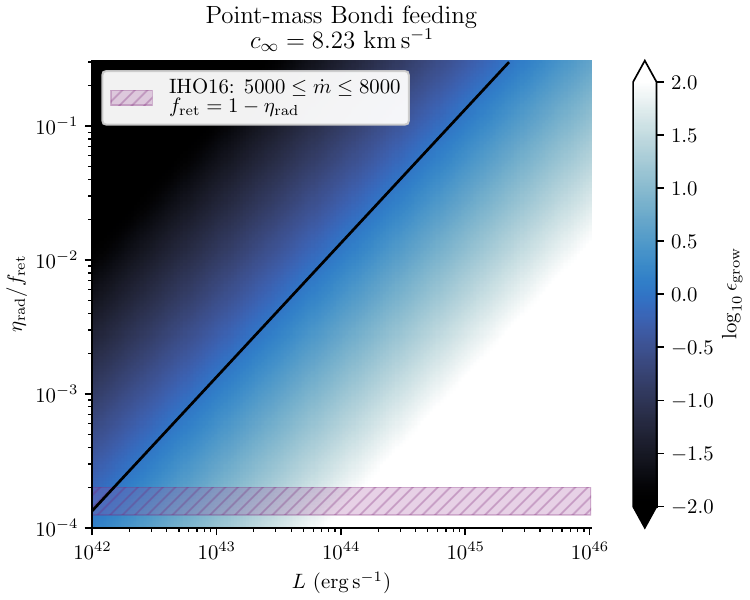}
\caption{Testing growth in a cold point-mass Bondi feeding model.
Color gives $\log_{10}\epsilon_{\rm grow}$, with
$\epsilon_{\rm grow}=t_{\rm B}/t_{\rm grow}$ the ratio of Bondi
sound-crossing to retained-growth time. Here $L$ is the escaping
accretion-powered luminosity; $\eta_{\rm rad}$ and $f_{\rm ret}$ are
the radiated and retained fractions of the same mass supply.
The sound speed
$c_\infty=8.23\ {\rm km\,s^{-1}}$ describes neutral atomic gas at
$10^4$ K with mean molecular weight $\mu=1.22$. The black contour
marks equal times; quasi-steady growth requires a ratio much smaller
than unity. The hatched band uses the IHO16 trapping prescription
\citep{2016MNRAS.459.3738I},
$\eta_{\rm rad}=3/(10+3\dot m)$ over $5000\leq\dot m\leq8000$,
with $\dot m=\dot M_{\rm flow}c^2/L_{\rm Edd}$ the supply in Eddington
units. Its endpoints span the published hyper-Eddington onset and worked
inflow rate. Retention is $f_{\rm ret}=1-\eta_{\rm rad}$, allowing only
escaping radiation to reduce growth; further losses move the band upward.
The Bondi clock requires point-mass dominance out to the Bondi radius
$R_{\rm B}$. Host-influenced flows use their own response time in
equation~(\ref{eq:flowresponsegrowth}).}
\label{fig:lrdlumgrowth}
\end{figure}

Let $t_{\rm resp}$ denote the response time of the part of the feeding flow
whose dynamics depends appreciably on the hole. The corresponding
growth comparison is
\begin{equation}
 \frac{t_{\rm resp}}{t_{\rm grow}}
 =\frac{L f_{\rm ret}t_{\rm resp}}
 {\eta_{\rm rad}M}.
 \label{eq:flowresponsegrowth}
\end{equation}
The mass cancels when the adopted response time scales linearly with
$M$, as it does for $t_{\rm B}$ at fixed sound speed. For a
general feeding model, $t_{\rm resp}$ must be obtained from its pressure,
velocity, and gravitational structure. The stellar velocity dispersion
$\sigma_\star$ gives an approximate influence scale
$R_{\rm infl}=M/\sigma_\star^2$. Using it to specify a gas
response time requires an additional dynamical assumption.

For the point-mass Bondi model, equation~(\ref{eq:growthloading}) compares the adjustment time to the
retained growth time:
\begin{equation}
 \epsilon_{\rm grow}=\frac{M/c_\infty^3}
 {M/\dot M}
 =\frac{\dot M}{c_\infty^3}.
\end{equation}
Substituting the luminosity and retained rate of that flow gives
\begin{equation}
 \epsilon_{\rm grow}
 =\frac{L}{c_\infty^3}\frac{f_{\rm ret}}{\eta_{\rm rad}}.
 \label{eq:lrdlumloading}
\end{equation}
A quasi-steady Bondi continuation requires
$\epsilon_{\rm grow}\ll1$. As the ratio approaches unity, the central
mass changes substantially during a Bondi adjustment time, and the growth
and flow must be evolved together.

For numerical evaluation we use the same atomic feeding medium as in
equation~(\ref{eq:growthnumber}). Its isothermal sound speed,
$c_\infty=(k_{\rm B}T_\infty/\mu m_p)^{1/2}$, is
$8.23\ {\rm km\,s^{-1}}$ after rounding. We choose
$10^{44}\ {\rm erg\,s^{-1}}$ as a rounded LRD luminosity reference:
the host-subtracted median stack of \citet{Sun2026aarie} has an inferred
bolometric luminosity $10^{43.9}\ {\rm erg\,s^{-1}}$. In applying the
growth criterion, we assign the adopted luminosity to the accretion-powered
component of the source.
Converting equation~(\ref{eq:lrdlumloading}) to these physical units gives
\begin{align}
 \epsilon_{\rm grow}&\simeq0.0133\left(\frac{L}{10^{44}\ {\rm erg\,s^{-1}}}\right)
 \frac{f_{\rm ret}}{\eta_{\rm rad}}\nonumber\\
 &\qquad\times\left(\frac{c_\infty}{8.23\ {\rm km\,s^{-1}}}\right)^{-3}.
\end{align}
Figure~\ref{fig:lrdlumgrowth} shows this ratio. Unit loading occurs at a retained growth
rate of $0.133\ M_\odot\,{\rm yr^{-1}}$, obtained by evaluating
$\dot M=c_\infty^3$ in physical units at this sound speed.
At $L=10^{44}\ {\rm erg\,s^{-1}}$, an interpretation with
$\eta_{\rm rad}/f_{\rm ret}<0.0133$ has a growth time shorter than
the Bondi crossing time. Doubling the feeding sound speed lowers the
crossover in $\eta_{\rm rad}/f_{\rm ret}$ by a factor of eight.
The mass independence holds at fixed luminosity, sound speed, and conversion
ratio. Along the similarity family itself, luminosity and growth loading
both increase with mass.

The feeding scale must also lie within the domain of the point-mass model.
At the adopted sound speed, restoring physical units gives
\begin{align}
 R_{\rm B}&\simeq635\ {\rm pc}
 \left(\frac{M}{10^7M_\odot}\right)
 \left(\frac{c_\infty}{8.23\ {\rm km\,s^{-1}}}\right)^{-2},\nonumber\\
 t_{\rm B}&\simeq75.4\ {\rm Myr}
 \left(\frac{M}{10^7M_\odot}\right)
 \left(\frac{c_\infty}{8.23\ {\rm km\,s^{-1}}}\right)^{-3}.
 \label{eq:lrdbondiscale}
\end{align}
The reference mass illustrates the upper end of the
$10^5$--$10^7M_\odot$ range considered by
\citet{2026Natur.649..574R}; at the $10^4M_\odot$ mass of the
\citetalias{2016MNRAS.459.3738I} example, these scales are
$0.635$ pc and $7.54\times10^4$ yr. Applying the cold Bondi clock at
LRD masses therefore requires the enclosed stellar, dark-matter, and gas
masses to remain small compared with $M$ throughout the feeding
region. The emitting size alone does not determine that mass budget.
Where the host contributes appreciably to gravity, its potential and the
gas dynamics set the feeding flow \citep{Hobbs2012oiilr}.

A published flow provides a worked example using its own luminosity
prescription. The trapping model of \citetalias{2016MNRAS.459.3738I}
sets $\eta_{\rm rad}=3/(10+3\dot m)$, with
$\dot m=\dot M_{\rm flow}/L_{\rm Edd}$ in our units. At the reported
$\dot m\simeq8000$, this gives $\eta_{\rm rad}\simeq1.25\times10^{-4}$
and $L\simeq L_{\rm Edd}$. For their $10^4M_\odot$ hole and the
$\kappa_{\rm es}=0.4\ {\rm cm^2\,g^{-1}}$ normalization,
the prescribed luminosity is $1.25\times10^{42}\ {\rm erg\,s^{-1}}$.
Using this luminosity and their $10^4$ K ambient medium in
equation~(\ref{eq:lrdlumloading}) gives
$\epsilon_{\rm grow}\simeq1.33f_{\rm ret}$, recovering the
rate-based result of equation~(\ref{eq:growthnumber}). A continuation that
retains nearly all the inflow therefore requires evolving the central mass
with the gas.

Figure~\ref{fig:lrdlumgrowth} extends this prescription across luminosity
for $5000\leq\dot m\leq8000$, from the approximate hyper-Eddington
onset to the reported inflow. Retaining all supplied mass-energy except
the escaping radiation gives $f_{\rm ret}=1-\eta_{\rm rad}$ and
$\eta_{\rm rad}/f_{\rm ret}\simeq(1.25$--$2.00)\times10^{-4}$.
At the LRD reference luminosity $10^{44}\ {\rm erg\,s^{-1}}$ and
the adopted atomic sound speed, this band gives
$\epsilon_{\rm grow}\simeq66.6$--$107$. Extending the trapping
prescription to that luminosity therefore implies growth much faster
than Bondi adjustment under the stated retention and feeding conditions.

To illustrate the sensitivity to the response model, choose
$\sigma_\star=100\ {\rm km\,s^{-1}}$ as a round comparison scale.
At $M=10^7M_\odot$, it gives $R_{\rm infl}=4.30$ pc.
If the gas responds on $t_{\rm resp}=R_{\rm infl}/\sigma_\star$,
the response time is $4.21\times10^4$ yr and the same luminosity and
conversion band give $t_{\rm resp}/t_{\rm grow}\simeq0.0371$--$0.0594$.
The reduction is $(8.23/100)^3\simeq5.57\times10^{-4}$.
These values evaluate an assumed crossing clock; the stellar dispersion
by itself supplies no measurement of the gas response. For example, the
acoustic crossing time at the same radius with the cold gas sound speed
is $5.11\times10^5$ yr. The pressure and inflow structure determine which
times govern adjustment. Thus the large values in
Figure~\ref{fig:lrdlumgrowth} test the specified cold Bondi continuation.

For an observed LRD, assigning the escaping accretion-powered luminosity
requires a bolometric and geometric conversion of the measured emission.
The sound speed describes the ambient feeding gas, whose temperature is
set outside the emitting photosphere. Radiative efficiency and retained
fraction must share an energy budget; absent an additional energy source,
$\eta_{\rm rad}+f_{\rm ret}\leq1$.

\section{Gas inventory, self-gravity, and column}\label{sec:inventory}

The same supply that powers the source must pass through a finite
gas reservoir. Steady continuity links its inventory to the transit time,
giving a second consistency test even when a small retained fraction
keeps black-hole growth slow.

\subsection{Growth loading and the mass of the feeding flow}\label{sec:gasmass}

Self-gravity adds another scale to the point-mass family. Define the gas loading
$\mathcal G=M_{\rm gas}(<R_{\rm B})/M$, where
$M_{\rm gas}(<R_{\rm B})$ is the gas mass enclosed by the Bondi radius. It scales as
$\mathcal G'=\lambda\mathcal G$ because the enclosed gas mass gains
$\lambda^2$.
Moving toward larger black-hole mass along fixed $n_\infty M$
therefore strengthens self-gravity. Cooling and radiative confinement trace
$n_\infty\propto M^{-1}$, whereas fixed gas loading traces
$n_\infty\propto M^{-2}$. Their intersection creates a mass scale even
though the point-mass equations contain none.

The family compares equivalent fixed-mass problems. Following the same curve
during growth requires the ambient density to
decline as $M^{-1}$ while all other reduced data remain fixed, and
even then equations~(\ref{eq:growthloading})--(\ref{eq:growthnumber}) show
where the quasi-steady interpretation reaches its boundary.

Gas moving slowly toward the hole spends longer in the feeding region.
Sustaining a given inward rate therefore places more gas in that region.
Steady spherical continuity expresses this mass budget directly. Let $r_{\rm in}$
be an absorbing inner boundary, take $v<0$, and suppose that no distributed
mass sources or sinks occur between $r_{\rm in}$ and $R_{\rm B}$. The inward
mass flux $\dot M_{\rm flow}=-4\pi r^2\rho v$ is then independent of radius.
Counting the gas outside the absorber, continuity gives the exact identity
\begin{equation}
 \begin{aligned}
 M_{\rm gas}(<R_{\rm B})
 &=4\pi\int_{r_{\rm in}}^{R_{\rm B}}\rho r^2\,dr
 =\dot M_{\rm flow}t_{\rm adv},\\
 t_{\rm adv}&\equiv\int_{r_{\rm in}}^{R_{\rm B}}\frac{dr}{|v(r)|}.
 \end{aligned}
 \label{eq:massresidence}
\end{equation}
Here $t_{\rm adv}$ is the time taken by gas to traverse the frozen profile. If
$f_{\rm ret}=\dot M/\dot M_{\rm flow}$, with
$0<f_{\rm ret}\leq1$, is the retained fraction at the inner boundary, then
\begin{equation}
 \mathcal G=\frac{\epsilon_{\rm grow}}{f_{\rm ret}}\,
 \chi,\qquad \chi\equiv\frac{t_{\rm adv}}{t_{\rm B}}.
 \label{eq:growthgasidentity}
\end{equation}
With distributed winds, the mass integral weights the local flux by the time spent at each
radius. A time-dependent reservoir additionally stores gas as supply and
drainage change.

Equation~(\ref{eq:growthgasidentity}) supplies a direct consistency test.
For a prescribed profile and retention, point-mass dominance requires
$\epsilon_{\rm grow}\chi/f_{\rm ret}\ll1$, in addition to the slow-growth
condition $\epsilon_{\rm grow}\ll1$. With a transit time of order $t_{\rm B}$,
sonic dilation brings the gas loading to order unity or larger.
An extended slow outer region already supplies a lower bound: if
$|v|\leq c_\infty$ throughout an outer
interval $r_s<r<R_{\rm B}$, then
$\chi\geq1-r_s/R_{\rm B}$. Small retention or a long transit
time allows a massive feeding region to accompany slow black-hole growth.

The critical isothermal Bondi solution gives a quantitative benchmark.
Writing $y=r/R_{\rm B}$ for scaled radius, define the inward Mach number
$\mathcal M_{\rm in}=|v|/c_\infty$. Its transonic branch obeys
\begin{equation}
 \mathcal M_{\rm in}^2-\ln \mathcal M_{\rm in}^2=4\ln(2y)+\frac{2}{y}-3,
 \label{eq:isothermalbenchmark}
\end{equation}
with $\mathcal M_{\rm in}>1$ for $y<1/2$ and $\mathcal M_{\rm in}<1$ for $y>1/2$
\citep{1952MNRAS.112..195B}. Numerical quadrature gives
\begin{equation}
 \chi=\int_0^1\frac{dy}{\mathcal M_{\rm in}(y)}\simeq1.05,\qquad
 \mathcal G\simeq1.05\,\frac{\epsilon_{\rm grow}}{f_{\rm ret}}.
 \label{eq:bondiresidence}
\end{equation}
The integral is finite at the origin because $\mathcal M_{\rm in}\sim(2/y)^{1/2}$.
Reducing the lower cutoff from $10^{-3}$ to $10^{-7}$ changes $\chi$ by
less than $1.5\times10^{-5}$. With full retention, $\mathcal G=1$ occurs
at $\epsilon_{\rm grow}\simeq0.953$, already near sonic dilation.
Gas takes about one Bondi time to reach the hole, so growth and gas gravity
become appreciable together in this profile.

\begin{figure}[t]
 \centering
 \includegraphics[width=\columnwidth]{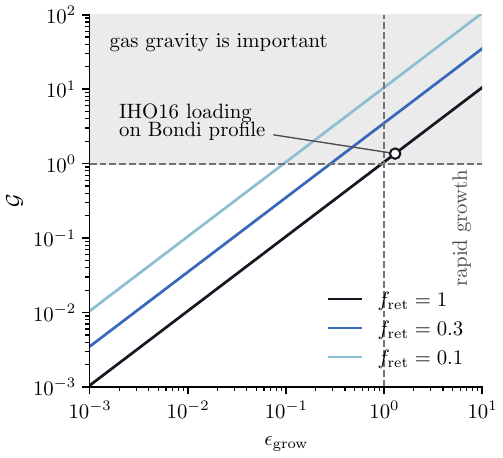}
 \caption{The gas mass required to sustain growth through the critical
 isothermal Bondi profile. The axes show retained growth loading
 $\epsilon_{\rm grow}=\dot M t_{\rm B}/M$ and enclosed gas
 loading $\mathcal G=M_{\rm gas}(<R_{\rm B})/M$.
 Gas spends approximately $1.05\,t_{\rm B}$ in transit, linking
 the mass of the feeding region to the rate retained by the hole. Lines show
 $\mathcal G\simeq1.05\epsilon_{\rm grow}/f_{\rm ret}$ for retained fractions
 $f_{\rm ret}=1,0.3,0.1$ at the inner boundary, chosen to illustrate full
 retention and two progressively smaller retained fractions over one decade.
 Gray dashed lines mark unit
 growth and gas loadings. The shaded region has $\mathcal G\geq1$, where
 gas contributes appreciably to gravity. Curves extrapolate the frozen
 test-fluid profile into that region. The open circle evaluates the
 isothermal profile at the rounded IHO16 growth loading $1.3$, with full
 retention.}
 \label{fig:growthgas}
\end{figure}

Applying the rounded $\epsilon_{\rm grow}=1.3$ to this benchmark gives
$\mathcal G\simeq1.36$ for full retention
(Figure~\ref{fig:growthgas}). Applying the same mass budget to the radiating
\citetalias{2016MNRAS.459.3738I} solution requires its density or velocity
profile to determine $\chi$. At a fixed feeding profile, lowering retention
reduces the hole's growth rate while preserving the mass of gas in transit.
Winds below the resolved domain can therefore accompany a slowly growing
hole inside a massive feeding region.

Along a similarity family with fixed retention, $\chi$ and
$\mathcal G/\epsilon_{\rm grow}$ are invariant, while both loadings
increase with the mass scale. Beyond point-mass dominance, the hole and
feeding gas jointly determine the potential and subsequent flow.

\subsection{Predicting the gas inventory and column}\label{sec:gascolumn}

The same mass budget connects luminosity to the gas required throughout
a specified feeding region, including one influenced by the host potential. Let $R_{\rm out}$ be the outer radius of the
specified feeding region. For steady spherical inflow with no distributed
mass sources or sinks, the same continuity argument as in
equation~(\ref{eq:massresidence}) gives
\begin{equation}
 \begin{aligned}
 t_{\rm tr}(R_{\rm out})&\equiv
 \int_{r_{\rm in}}^{R_{\rm out}}\frac{dr}{|v(r)|},\\
 M_{\rm gas}(<R_{\rm out})
 &=\dot M_{\rm flow}t_{\rm tr}(R_{\rm out})
 =\frac{L t_{\rm tr}(R_{\rm out})}{\eta_{\rm rad}}.
 \end{aligned}
 \label{eq:generalinventory}
\end{equation}
Here $t_{\rm tr}$ is the time gas spends in transit through the specified
radial interval, and $t_{\rm tr}(R_{\rm B})=t_{\rm adv}$. The inventory
relation also holds when the host or gas contributes to gravity, provided
the velocity profile describes the adopted steady flow. At fixed luminosity,
radiative efficiency, and transit time, the required gas mass is independent
of retention at the inner boundary. Reducing retention slows the hole's
growth while leaving this inventory unchanged. The transit time controls
the inventory; the response time $t_{\rm resp}$ controls the growth test.
Their relation follows from the flow dynamics.

The required reservoir can also be tested against the total dynamical
mass enclosed within $R_{\rm out}$. The gas and central hole must fit
within that budget, with stellar and dark-matter contributions reducing
the mass available to gas. The predicted inventory and the dynamical
constraint must refer to the same feeding radius.

The inventory also places a lower bound on the radial mass column through
the same spherical feeding region. Define
\begin{equation}
 \Sigma_{\rm rad}(R_{\rm out})\equiv
 \int_{r_{\rm in}}^{R_{\rm out}}\rho(r)\,dr.
 \label{eq:radialmasscolumn}
\end{equation}
Every shell lies at $r\leq R_{\rm out}$, so the mass integral obeys
\begin{equation}
 \begin{aligned}
 M_{\rm gas}(<R_{\rm out})
 &=4\pi\int_{r_{\rm in}}^{R_{\rm out}}\rho r^2\,dr\\
 &\leq4\pi R_{\rm out}^2\Sigma_{\rm rad}(R_{\rm out}).
 \end{aligned}
\end{equation}
Substituting the inventory gives
\begin{equation}
 \Sigma_{\rm rad}(R_{\rm out})\geq
 \frac{L t_{\rm tr}(R_{\rm out})}
 {4\pi\eta_{\rm rad}R_{\rm out}^2}.
 \label{eq:generalcolumnbound}
\end{equation}
The geometric inequality holds for any nonnegative spherical density
profile; its luminosity form uses the steady mass budget. Placing a fixed
mass farther inward increases its column. Thus the adopted feeding radius
and velocity profile give a minimum column even when the gravitational
potential contains several components.

Taking $R_{\rm out}=R_{\rm B}$ recovers the Bondi-scale gas loading:
\begin{equation}
 \mathcal G=\epsilon_{\rm grow}\frac{\chi}{f_{\rm ret}}
 =\frac{L\chi}{\eta_{\rm rad}c_\infty^3},\qquad
 \chi=\frac{t_{\rm adv}}{t_{\rm B}}.
 \label{eq:lumgasinventory}
\end{equation}
Using $R_{\rm B}=M/c_\infty^2$, the column bound becomes
\begin{equation}
 \Sigma_{\rm rad}(R_{\rm B})\geq
 \frac{\mathcal G M}{4\pi R_{\rm B}^2}
 =\frac{L\chi c_\infty}{4\pi\eta_{\rm rad}M}.
 \label{eq:lumcolumnbound}
\end{equation}
Along an atomic similarity family, $L/M$ remains fixed, so this
bound respects the preserved columns. The critical isothermal Bondi
profile supplies the benchmark $\chi\simeq1.05$. As its inferred
$\mathcal G$ approaches unity, gas gravity must enter the calculation
of the velocity profile and hence of $\chi$.

A radial profile sharpens the bound into a column prediction. For example,
extending the prescribed $\rho=Ar^{-3/2}$ atmosphere from $r_{\rm in}$
to $R_{\rm out}$ gives
\begin{equation}
 \begin{aligned}
 M_{\rm gas}(<R_{\rm out})
 &=\frac{8\pi A}{3}
 \left(R_{\rm out}^{3/2}-r_{\rm in}^{3/2}\right),\\
 \Sigma_{\rm rad}(R_{\rm out})
 &=2A\left(r_{\rm in}^{-1/2}-R_{\rm out}^{-1/2}\right).
 \end{aligned}
 \label{eq:prescribedcolumn}
\end{equation}
When $r_{\rm in}\ll R_{\rm out}$, the outer gas dominates the mass and
the inner gas dominates the column. Connecting the compact emitting
atmosphere to the feeding reservoir therefore requires its radial extent.

For a fixed composition, define the total column of hydrogen nuclei by
$N_{\rm H}(R_{\rm out})=\Sigma_{\rm rad}(R_{\rm out})/(\mu_{\rm H}m_p)$,
where $\mu_{\rm H}m_p$ is the mass per hydrogen nucleus. The numerical
strength of the Bondi-scale bound can be illustrated with a hydrogen mass
fraction $0.76$, consistent with the neutral primordial composition used
for $c_\infty$, giving $\mu_{\rm H}=1/0.76$. In physical units,
\begin{align}
 N_{\rm H}(R_{\rm B})&\gtrsim2.62\times10^{19}\ {\rm cm^{-2}}
 \left(\frac{L}{10^{44}\ {\rm erg\,s^{-1}}}\right)
 \left(\frac{\chi}{1.05}\right)\nonumber\\
 &\quad\times\left(\frac{\eta_{\rm rad}}{0.1}\right)^{-1}
 \left(\frac{M}{10^7M_\odot}\right)^{-1}\nonumber\\
 &\quad\times\left(\frac{c_\infty}{8.23\ {\rm km\,s^{-1}}}\right)
 \left(\frac{\mu_{\rm H}}{1/0.76}\right)^{-1}.
 \label{eq:columnnormalization}
\end{align}
The reference efficiency $0.1$ illustrates efficient radiation; choosing
$10^{-4}$ illustrates the low efficiencies near the trapping band.
At the other reference values, the latter gives a formal minimum
$2.62\times10^{22}\ {\rm cm^{-2}}$, increasing to
$2.62\times10^{23}\ {\rm cm^{-2}}$ at $10^6M_\odot$.
The same assumptions give $\mathcal G\simeq0.140$ for efficiency $0.1$
and $\mathcal G\simeq1.40\times10^2$ for $10^{-4}$, independent of mass.
For $10^6M_\odot$ and $10^7M_\odot$ holes, the latter corresponds
to required reservoirs of $1.40\times10^8M_\odot$ and
$1.40\times10^9M_\odot$, respectively.
The low-efficiency numbers therefore extrapolate a test-fluid profile
deep into the self-gravitating regime. They quantify the inventory demanded
by that assumed flow. A physical column prediction there requires its
revised transit time and feeding radius in
equation~(\ref{eq:generalcolumnbound}).

Balmer absorption and electron-scattering wings offer observational access
to envelope columns and ionization
\citep{2026Natur.649..574R,Matthee2026dtfpl}.
For example, \citet{2026Natur.649..574R} infer electron columns of
approximately $(0.7$--$4.2)\times10^{24}\ {\rm cm^{-2}}$.
Comparing the predicted hydrogen column with a spectrum requires the
ionization and level populations, the emitting
radius, and the sightline's covering geometry. In particular, photons
emitted within the reservoir sample the gas outside their emitting
surface. Equation~(\ref{eq:generalcolumnbound}) refers to the full radial
interval from $r_{\rm in}$ to $R_{\rm out}$. For a proposed spherical
model, an independent upper limit on that full column below the predicted
minimum would exclude its adopted luminosity, efficiency, transit time,
and feeding radius together. A column above the minimum can be tested against the
profile prediction in equation~(\ref{eq:prescribedcolumn}) or its analogue
for the chosen flow. This provides a direct comparison between the gas
required to sustain the luminosity and the gas inferred from the spectrum.

\section{Concluding Remarks}\label{sec:conclusions}

Atomic accretion onto a point mass admits an exact similarity at fixed
ambient temperature and composition. Increasing the black-hole mass $M$,
radii, and times while decreasing density by the same factor preserves
the dimensionless flow when two-body reaction rates, transport, radiation,
and boundary conditions obey the required scalings. The product
$n_\infty M$, with $n_\infty$ the ambient number density, labels a family
once the remaining dimensionless data are fixed. This transformation
connects simulations across mass scales and explains the common
mass--density slope of radiative accretion boundaries.

For absorption proportional to density and emission proportional to
density squared at fixed temperature and composition, the similarity
preserves normalized spectra, equivalent widths, and velocity profiles.
Density-sensitive atomic populations and thermal emission in local
thermodynamic equilibrium break these scalings. Public atmosphere models
of Little Red Dots illustrate the resulting density information:
rest-frame near-infrared coverage distinguishes emitting densities that
optical data leave degenerate, with sensitivity to contamination and
extinction. Even at known luminosity and density profile, a prescribed
spherical inflow retains a degeneracy between central mass and inward
drift. Inferring mass requires an additional dynamical constraint.
This freedom holds the radiating atmosphere fixed; along the atomic
similarity family, luminosity scales with mass and can set the mass
scale when their ratio is calibrated.

Growth introduces a clock that breaks the fixed-mass similarity.
In geometrized units ($G=c=1$), gas with ambient sound speed $c_\infty$
has Bondi radius $R_{\rm B}=M/c_\infty^2$ and crossing time
$t_{\rm B}=R_{\rm B}/c_\infty$. The retained mass rate $\dot M$ sets the
growth time $t_{\rm grow}=M/\dot M$. For point-mass Bondi feeding,
\[
 \epsilon_{\rm grow}\equiv\frac{t_{\rm B}}{t_{\rm grow}}
 =\frac{L f_{\rm ret}}{\eta_{\rm rad}c_\infty^3},
\]
where $L$ is the escaping accretion-powered luminosity, $\eta_{\rm rad}$
is the radiated fraction of the supplied rest-mass energy, and
$f_{\rm ret}$ is the retained fraction of the same supply.
At fixed sound speed, this ratio also measures the expansion speed of
$R_{\rm B}$ relative to $c_\infty$. Quasi-steady growth requires
$\epsilon_{\rm grow}\ll1$; near unity, the central mass and feeding gas
must evolve together. At fixed luminosity and conversion factors, this
test is independent of the assigned mass but depends strongly on the
feeding sound speed. It requires point-mass dominance throughout the
Bondi region; a host-influenced flow must supply its own response time.
A retained growth law sustained indefinitely as a positive power of mass
also leaves only finitely many future Bondi crossing times.

The feeding region must contain enough gas to sustain the supply.
In steady spherical flow without distributed mass sources or sinks,
the gas inventory equals the inward mass flux multiplied by its transit
time. For the critical isothermal Bondi profile, transit from $R_{\rm B}$
to a small central absorber takes about $1.05\,t_{\rm B}$. With full
retention, $\epsilon_{\rm grow}\simeq1$ therefore requires gas mass
comparable to the hole's mass, making gas self-gravity appreciable.

More generally, let $R_{\rm out}$ bound a specified feeding region and
$t_{\rm tr}$ be its transit time to the inner absorber. The same steady
mass budget requires gas mass $M_{\rm gas}=L t_{\rm tr}/\eta_{\rm rad}$
and radial mass column
$\Sigma_{\rm rad}\geq M_{\rm gas}/(4\pi R_{\rm out}^2)$.
Lower retention slows growth while leaving this inventory unchanged at
fixed luminosity, efficiency, and transit time. Comparison with absorption
and scattering measurements requires the ionization and emitting geometry
to relate the observed column to that same radial interval.
The spectrum, dynamics, and gas inventory thus provide complementary
tests of an accretion interpretation: the emitting gas must belong to a
flow that can supply the luminosity and respond to the growing hole.

\begin{acknowledgments}
I thank Christopher Irwin for pointing out the illuminating work of
\citet{2012ApJ...754..154M}, which inspired this work; Jenny Greene and her
group for helpful discussions and for patiently answering my elementary
questions during group meetings; and Eliot Quataert for general discussions
surrounding accretion physics.
\end{acknowledgments}

\software{\href{https://libraries.io/pypi/astrobib}{\texttt{astrobib}},
\texttt{CMasher} \citep{vanderVelden2020myawj},
\texttt{Matplotlib} \citep{2007CSE.....9...90H},
\texttt{NumPy} \citep{Harris2020qkmow}}

\clearpage
\appendix
\renewcommand{\theHequation}{\thesection.\arabic{equation}}
\section{Accumulated growth clocks and growing profiles}\label{sec:growingderivations}
We consider a spherical flow around a black hole of mass $M(t)$
whose retained growth rate is $\dot M>0$. The ambient sound speed
$c_\infty$ is fixed. In geometrized units $G=c=1$, the instantaneous
Bondi radius and time are $R_{\rm B}=M/c_\infty^2$ and
$t_{\rm B}=M/c_\infty^3$. The growth time is
$t_{\rm grow}=M/\dot M$, and the growth loading is
$\epsilon_{\rm grow}=t_{\rm B}/t_{\rm grow}$. These definitions specify
the clock calculation. The stationary-profile argument additionally
assumes the atomic similarity scalings, fixed reduced boundary data,
and a fixed retained fraction of the inward flux.

\subsection{Accumulated Bondi time}
Introduce the
accumulated number of instantaneous Bondi times,
\begin{equation}
 d\mathcal N\equiv\frac{dt}{t_{\rm B}(t)}.
 \label{eq:growingbonditime}
\end{equation}
Since $t_{\rm B}/M=c_\infty^{-3}$ is constant,
\begin{equation}
 d\mathcal N=\frac{d\ln M}{\epsilon_{\rm grow}(M)}.
 \label{eq:growingclockidentity}
\end{equation}
For any unbounded retained growth history beginning at time $t_0$ with
$M_0=M(t_0)$, the future flow time is
\begin{equation}
 \Delta\mathcal N_\infty=c_\infty^3
 \int_{t_0}^{t_\star}\frac{dt}{M(t)}
 =c_\infty^3
 \int_{M_0}^{\infty}\frac{dM}{M\dot M}.
 \label{eq:generalflowhorizon}
\end{equation}
Here $t_\star$ is the endpoint at which the idealized history reaches
unbounded mass; it may be finite, as for a sustained power law whose growth
exponent $p$ exceeds unity,
or infinite. The flow-time horizon is finite exactly when either integral converges.
For a power-law episode, let
$\epsilon_0\equiv\epsilon_{\rm grow}(M_0)$ and
$\dot M\propto M^p$ with $p>0$. Then
$\epsilon_{\rm grow}=\epsilon_0(M/M_0)^p$, and integration to
arbitrarily large mass gives the maximum additional accumulated Bondi time,
\begin{equation}
 \Delta\mathcal N_{\max}=\frac{1}{p\epsilon_0}.
 \label{eq:bonditimehorizon}
\end{equation}
Thus every track that sustains this power law has a finite flow-time horizon.
Equation~(\ref{eq:generalflowhorizon}) also covers non-algebraic histories.
A sustained power law with $p>0$ beginning at
$\epsilon_{\rm grow}<1$ reaches the sonic dilation
boundary after a finite increase in mass. For a constant retained
Eddington-scaled rate ($p=1$), this occurs at
$M/M_0=1/\epsilon_0$. Starting instead at the rounded IHO16 loading
$\epsilon_0=1.3$, the same retained law accumulates at most $0.769$
additional Bondi times. Retaining a fixed fraction of the Bondi
supply in a fixed ambient medium gives
$\dot M_{\rm B}\propto\rho_\infty M^2$, $p=2$, and at most $0.385$
additional Bondi times. The same formal $M^2$ runaway appears in
early-Universe accretion estimates and supra-exponential seed-growth scenarios
\citep{1967SvA....10..602Z,Alexander2014tivle}. Zel'dovich and Novikov also
noted the eventual loss of its stationary-flux assumption; equation
(\ref{eq:bonditimehorizon}) expresses that loss in the instantaneous flow
clock. At $\epsilon_0=1.3$, both idealized laws therefore supply less
than one subsequent global flow-crossing time even if continued indefinitely. A finite
reservoir, self-gravity, feedback, or environmental evolution will intervene
well before the formal infinite-mass limit used in deriving equation
(\ref{eq:bonditimehorizon}), and before a succession of globally relaxed
fixed-mass states can be constructed.
For a finite mass increase $f_M=M/M_0$, the same integration gives
\begin{equation}
 \Delta\mathcal N(f_M)=\frac{1-f_M^{-p}}{p\epsilon_0}.
 \label{eq:finitemassclock}
\end{equation}
At $\epsilon_0=1.3$, even a mass doubling accumulates only $0.385$ Bondi
times for $p=1$ and $0.288$ for $p=2$. These are conditional clock budgets
for the assumed retained laws. They measure the amount of global flow time
available while the central mass increases.

\subsection{The stationary growing-profile obstruction}
The obstruction can be seen directly in coordinates that follow
the growing capture radius. Let
\begin{equation}
 y=\frac{r}{R_{\rm B}(t)},\qquad
 a\equiv\frac{d\ln R_{\rm B}}{d\mathcal N}=\epsilon_{\rm grow},
 \label{eq:growingcoordinates}
\end{equation}
Here $r$ is spherical radius. Let $\rho$ denote gas density and $v$ radial
velocity, and write $\rho=\rho_0(t)D(y,\mathcal N)$,
$v=c_\infty V(y,\mathcal N)$, and
$\beta_\rho=d\ln\rho_0/d\mathcal N$, where $\rho_0(t)$ is the density scale and $D$ and $V$
are the reduced density and radial velocity. In spherical symmetry the continuity equation becomes
\begin{equation}
 \partial_{\mathcal N}D-ay\partial_yD+\beta_\rho D
 +\frac{1}{y^2}\partial_y(y^2DV)=0.
 \label{eq:growingcontinuity}
\end{equation}
The radial momentum equation contains the corresponding dilation term,
\begin{equation}
 \partial_{\mathcal N}V+(V-ay)\partial_yV
 =-\frac{1}{D}\partial_y\Pi-\frac{1}{y^2}+\mathcal F_{\rm rad},
 \label{eq:growingmomentum}
\end{equation}
where $\Pi=P/(\rho_0c_\infty^2)$ is reduced gas pressure and
$\mathcal F_{\rm rad}$ denotes the dimensionless radiative acceleration.
Energy, chemistry, and transfer acquire analogous terms. For a constant
retained Eddington-scaled rate $\dot m_{\rm ret}=\dot M/L_{\rm Edd}$,
where $L_{\rm Edd}$ is the Eddington luminosity,
$da/d\mathcal N=a^2$, so the reduced equations are nonautonomous and $a$ diverges
at the finite flow-time horizon.

A nontrivial stationary accretion profile in the growing coordinates must
have fixed dimensionless parameters, including the dilation rate $a$.
Atomic similarity, however, requires $\rho_0\propto M^{-1}$; at fixed
nonzero reduced mass flux this gives
$\dot M\sim\rho_0R_{\rm B}^2c_\infty\propto M$ and hence
$a\propto M$. Conversely, holding $a$ fixed requires
$\dot M={\rm constant}$ and therefore
$\rho_0\propto M^{-2}$, which is incompatible with matching the flow
and two-body cooling scalings. Stationarity while $a$ changes requires its coefficients in
equations~(\ref{eq:growingcontinuity}) and (\ref{eq:growingmomentum}) to
vanish. With $\beta_\rho=-a$, these conditions and stationary mass conservation are
\begin{equation}
 y\partial_yD+D=0,\quad
 y\partial_yV=0,\quad
 \partial_y(y^2DV)=0.
 \label{eq:stationaryobstruction}
\end{equation}
The first two give $D\propto y^{-1}$ and constant $V$, for which $y^2DV$
is constant only when the flux vanishes. Thus no
nontrivial stationary growing profile preserves both the original atomic
similarity and its self-consistent retained flux. A time-dependent
solution remains possible and requires a new reduced profile beyond the
substitution $M\mapsto M(t)$.

This Newtonian obstruction echoes the result of \citet{1974MNRAS.168..399C}
that a primordial black hole cannot grow apace with an exact
radiation-dominated Friedmann exterior through self-similar accretion. A
classical constant-$a$ solution appears once self-gravity replaces the uniform
ambient medium: the inside-out collapse of a singular isothermal sphere has
$\dot M=0.975c_\infty^3$, hence $a=0.975$, an expansion wave moving at
$c_\infty$, and the required $\rho_0\propto M^{-2}$ scaling
\citep{1977ApJ...214..488S}. It lies close to the sonic boundary.

\section{Public spectral comparisons and numerical controls}\label{sec:lrdmethods}

The comparisons below support two steps in Section~\ref{sec:lrdfeeding}:
measuring departures along the spectral density--luminosity tracks and
determining how well the adopted atmosphere library distinguishes
photospheric densities. The synthetic targets isolate wavelength coverage
and contaminating components under controlled assumptions. Their recovery
supplies the density-information step of the argument; the remaining
mass--drift freedom follows from the prescribed flow relations.

\subsection{Spherical spectral residuals}
We use all 15 spectra from the public release accompanying
\citet{Liu2025drrkn}.\footnote{\url{https://doi.org/10.5281/zenodo.17204644}}
The models prescribe $\rho=A r^{-3/2}$ and solve radiation and temperature
in a prescribed density profile. At $L_0=1.47\times10^{44}\ {\rm erg\,s^{-1}}$,
the five tracks have density at $R_0=1.477\times10^{16}\ {\rm cm}$ with
$\log_{10}[\rho(R_0)/({\rm g\,cm^{-3}})]=-13,-12.5,-12,-11.5,-11$.
Each track also contains $L_0/10$ and $10L_0$, with $A\propto L^{1/2}$.
The numerical reference radius, luminosity levels, and density samples
come from the released grid. Grouping them into these five tracks uses
every released spectrum and supplies a common central comparison model
for each track.
The transformed power law has $A'=\lambda^{1/2}A$, since
$\rho'(r')=\lambda^{-1}A(r'/\lambda)^{-3/2}$.
These tracks test the density and luminosity weights. Boundary radii and
thermal structures enter separately in interpreting the residuals.

For each endpoint, we form the magnitude difference relative to its
$L_0$ spectrum. Over bins wholly contained in 4000--10,000 \AA, we fit
$b_0+b_1(\lambda_{\rm wav}/5500\,\text{\AA})^{-1.2}$ by least squares,
weighted by bin width in $\ln\lambda_{\rm wav}$. The fitted
coefficients $b_0$ and $b_1$ set the offset and tilt. Here
$\lambda_{\rm wav}$ denotes wavelength. The unrestricted coefficient
$b_1$ allows the fit to remove a smooth tilt of either sign. Reported residuals are weighted
root-mean-square magnitude differences. The first four tracks give
0.010--0.019 mag at their endpoints; the densest gives 0.049 and 0.093 mag.
Including 3000--4000 \AA\ raises the largest residual to 0.325 mag.
The primary interval samples the optical continuum to the red of the
Balmer edge; extending it to 3000 \AA\ includes the break region. The
5500 \AA\ pivot places the tilt near the middle of the optical range,
and the exponent $-1.2$ supplies a simple, smooth wavelength dependence.
Both this tilt and the screen below are prescribed comparison functions.
The stored wavelength values are treated as \AA, as verified by their
product with the independent frequency array, despite the file's cm label.

\subsection{Atmosphere templates and mock recovery}
The model-statistics table of \citet{Liu2026mbxlp} contains 255 entries,
including 242 with radiation force. We use the complete table and 11
selected high-resolution spectra from the authors' CC BY 4.0
library, pinned to Git revision
\texttt{1d996420c61a62844ae50320e6fa7dbfc733596e}.\footnote{\url{https://github.com/hanpu-liu/LRD_synthetic_spectral_library}}
Downloaded files are checked against Git blob hashes and recorded with
SHA256 hashes. Table-derived results use the authors' seven continuum
bands, denoted here by $\widetilde B,\widetilde R_1,\widetilde R_2,\widetilde Z,\widetilde J,\widetilde H,\widetilde K$, with wavelength intervals in
\AA\ of 4500--4750, 5950--6200, 6750--7000, 8700--9000,
11700--12300, 15600--16400, and the union of 20800--21550 with
21750--21900, respectively. Tildes distinguish band labels from physical variables.
These custom continuum windows follow the
library's band definitions. Independent photon-counting integrations use
$\int F_\nu\,d\ln\lambda_{\rm wav}/\int d\ln\lambda_{\rm wav}$ and
recover the table's relative magnitudes within $5.1\times10^{-5}$ mag.
The 11 full spectra provide temperature, gravity, metallicity, and
microturbulence comparisons around the selected targets; the full
model-statistics table supplies the broader template search.

Five supplied atmospheres are noiseless targets: $T_{\rm eff}=4500$ K
at $\log g=-3,-2,-1$, and $T_{\rm eff}=4000,5000$ K at $\log g=-2$,
all with metallicity $[M/H]=-1$ and microturbulence
$2\ {\rm km\,s^{-1}}$. Every target is compared with all 242
radiation-force-on templates, allowing their available temperature,
metallicity, and microturbulence variations at the supplied grid nodes.
The 4500 K sequence samples three gravity values at fixed temperature;
the 4000 and 5000 K targets test the adjacent temperature range at a common
gravity. Fixing the target metallicity and microturbulence isolates these
comparisons while allowing both quantities to vary among the competing
templates. The choices are illustrative nodes from the supplied library.
Three coverage choices use the first four bands, the first five, or all
seven. Adding $\widetilde J$ extends the continuum baseline; adding
$\widetilde H$ and $\widetilde K$
samples across the H$^-$ feature near $1.6\,\mu$m. Two emission models
give $5\times3\times2=30$ experiments in total. The clean model
fits a free atmosphere amplitude. The mixed model additionally fits
nonnegative amplitudes of a flat-$F_\nu$ host template and a 1000 K Planck
dust template, with a screen applied to the atmosphere alone:
\begin{equation}
 A_{\lambda_{\rm wav}}=A_V
 (\lambda_{\rm wav}/5500\,\text{\AA})^{-1.2}.
\end{equation}
The screen is evaluated at band centers, scanning $0\leq A_V\leq0.6$
in steps of 0.01. At each screen value we choose nonnegative component
amplitudes to minimize the weighted sum of squared flux residuals, using
nonnegative least squares \citep{Lawson1974rrzuc}.
The mixed targets inject $A_V=0.2$, host flux equal to 0.3 times the
unattenuated atmosphere's $\widetilde R_1$ flux, and dust flux equal to 0.2 times its
unattenuated $\widetilde K$ flux at $\widetilde K$. The host spectrum and screen are controlled
nuisance templates chosen to explore how smooth contaminating components
affect the recovery of atmospheric density.
The flat host component supplies a smooth optical contribution. At
1000 K the Planck component rises through the near-infrared bands, so it
tests how a warm dust contribution affects the same curvature used for
density inference. The injected fractions of 0.3 and 0.2 keep both
components appreciable while the atmosphere remains prominent at their
normalization bands. The injected $A_V=0.2$ lies inside the search interval
0--0.6. Together these values define one reproducible contamination
scenario. Its temperature, fractions, and screen exponent are held fixed
in this experiment; a source-specific analysis would fit a broader range
of component shapes. The 0.01 screen step is checked by a fourfold refinement.

Differentiating the magnitude definition $m=-2.5\log_{10}F+{\rm constant}$
gives $|dm|=(2.5/\ln10)|dF|/F$. Each band therefore has independent assigned error
$\sigma_F/F=(\ln10/2.5)\sigma_m$, with $\sigma_m=0.05$ mag in the
primary comparisons. For each template and choice of nuisance parameters
we evaluate
\begin{equation}
 \mathcal Q=\sum_{b\in\mathcal B}
 \left(\frac{F_b^{\rm target}-F_b^{\rm model}}{\sigma_b}\right)^2,
 \qquad \Delta\mathcal Q=\mathcal Q_{\rm prof}-\mathcal Q_{\min}.
\end{equation}
The adopted 0.05-mag scale corresponds to about 4.61 percent fractional
flux error, or signal-to-noise ratio 21.7 per band. It sets a common precision
for comparing wavelength coverage. The 0.02- and 0.10-mag controls below
span approximately 1.84--9.21 percent errors.
Here $\mathcal B$ is the chosen set of bands, $\mathcal Q_{\rm prof}$ is the
minimum over the component amplitudes and screen for that atmosphere,
and $\mathcal Q_{\min}$ is the lowest value across all atmosphere templates.
We retain discrete templates with
$\Delta\mathcal Q\leq4$ relative to the minimum and report their density
extrema. This threshold defines the set of supplied templates compatible
with each synthetic target at the assigned error scale. Its width measures
discrimination within the discrete library. Sparse sampling and grid edges
can artificially narrow the set; interpreting such widths as observational
uncertainties would require a statistical treatment of real data and model
systematics.
For the lowest-density 4500 K target,
$\log_{10}[\rho_{\rm ph}/({\rm g\,cm^{-3}})]\simeq-11.14$, the primary
density width is 3.24 dex with optical coverage and 0.27 dex with JHK
added for the atmosphere alone. Including the specified host, dust, and
extinction gives widths of 4.51 and 1.27 dex, respectively.
We adopt four as a common squared-residual tolerance for displaying these
sets and also evaluate thresholds of one and nine. For the lowest-density
4500 K target with nuisance components, the optical density widths are
4.25, 4.51, and 4.51 dex at thresholds one, four, and nine, respectively;
with all seven bands they are 0.53, 1.27, and 1.74 dex. Near-infrared
coverage improves discrimination across these choices, while the quoted
width remains sensitive to the adopted tolerance.
For the lowest-density 4500 K target with all seven bands and nuisance
components, changing $\sigma_m$ to 0.02 and 0.10 mag gives widths of
0.53 and 2.53 dex, respectively. Including the 13 radiation-force-off
templates changes the primary 1.27-dex width to 1.32 dex. Refining the
screen step to 0.0025 leaves the 35 accepted primary templates unchanged.
These checks measure sensitivity to the assigned errors, radiation-force
treatment, and screen-search spacing. Additional atmosphere physics and
observational systematics would enter an application to measured spectra.

\subsection{Analytic and implementation checks}
Six focused tests verify formal transfer against an independent ODE
integration, two-level statistical equilibrium and its limiting regimes,
both null directions of equation~(\ref{eq:lrdjacobian}), optical-depth
consistency of a local opacity scaling, recovery of injected nuisance
components and rejection of an incorrect spectral shape, and mass
cancellation in equation~(\ref{eq:lrdlumloading}). All pass. The comparisons
measure the information available within prescribed public spectral models.
They provide a basis for applying the same questions to observed LRDs and
to models that evolve radiation and feeding dynamics together.

\bibliography{refs}
\bibliographystyle{aasjournalv7}
\end{document}